\documentclass[useAMS,usenatbib]{mn2e}
\usepackage{epsfig}
\usepackage{amsmath}

\def\lsim{~\rlap{$<$}{\lower 1.0ex\hbox{$\sim$}}}

\def\gsim{~\rlap{$>$}{\lower 1.0ex\hbox{$\sim$}}}

\def \HII{H\,{\sevensize II}\,\,}

\def \HIeqn{\rm H\,{\scriptscriptstyle I}}

\def \xvec{\textbf{\textit{x}}}

\def \yvec{\textbf{\textit{y}}}

\def \zvec{\textbf{\textit{z}}}

\def \kvec{\textbf{\textit{k}}}

\def \Uvec{\textbf{\textit{U}}}

\def \Msolar{M_{\odot}}

\title[Quasars \& the 21-cm power spectrum]{Modification of the 21-cm power spectrum by quasars during the epoch of reionisation}

\author[Geil \& Wyithe]{Paul M. Geil\thanks{Email: pmgeil@unimelb.edu.au}, J. Stuart B. Wyithe\\
School of Physics, University of Melbourne, Parkville, Victoria,
Australia}

\begin{document}

\maketitle

\label{firstpage}
\begin{abstract}
We assess the effect of a population of high-redshift quasars on the 21-cm power spectrum during the epoch of reionisation. Our approach is to implement a semi-numerical scheme to calculate the three-dimensional structure of ionised regions surrounding massive halos at high redshift. We include the ionising influence of luminous quasars by populating a simulated overdensity field with quasars using a Monte Carlo Markov Chain algorithm. We find that quasars modify both the amplitude and shape of the power spectrum at a level which is of the same order as the fractional contribution to reionisation. The modification is found both at constant redshift and at constant global neutral fraction, and arises because ionising photons produced by quasars are biased relative to the density field at a level that is higher than steller ionising photons. Our results imply that quasar ionisation will need to be included in detailed modelling of observed 21-cm power spectra.
\end{abstract}

\begin{keywords}
cosmology: diffuse radiation, theory--galaxies: high redshift, intergalactic medium
\end{keywords}

% ------------------------------------------------------------------------------------------------------------------------------------
\section{Introduction}
\label{Introduction}
The reionisation of cosmic hydrogen is commonly believed to have been due to UV photons produced by the first stars and quasars and was an important milestone in the history of the Universe \citep[e.g.,][]{bl2001}. Reionisation starts with isolated regions of ionised hydrogen (\HII) forming around galaxies and quasars, which later grow and merge to surround clusters of galaxies. The reionisation process is complete when these \HII regions overlap and fill the volume between galaxies. Various experiments are currently underway to measure 21-cm emission from the pre-overlap intergalactic medium (IGM) and thus observe the evolution of the ionisation stucture directly. These experiments include the Low Frequency Array\footnote{http://www.lofar.org/} [LOFAR], the Murchison Widefield Array\footnote{http://www.haystack.mit.edu/ast/arrays/mwa/} (MWA) and the Precision Array to Probe Epoch of Reionization\footnote{http://astro.berkeley.edu/$\sim$dbacker/eor/} (PAPER).

Several probes of the reionisation epoch in redshifted 21-cm emission have been suggested, which include: observation of emission as a function of redshift averaged over a large area of sky; imaging of individual \HII regions; observation of the power spectrum of fluctuations. Observing emission as a function of redshift averaged over a large area of sky provides a direct probe of the evolution in the neutral fraction of the IGM, and is referred to as the global step \citep{shaver1999,gnedin2004,furl2006a}. Observation of individual \HII regions in the image regime will probe quasar physics as well as the evolution of the neutral gas \citep{wl2004b,kohler2005,geil2008}. The most promising probe is the observation of the power spectrum of fluctuations together with its evolution in redshift \citep[see, e.g.,][]{furl2006b}. This observation would trace the evolution of neutral gas with redshift as well as the topology of the reionisation process via the spatial dependence of the statistics of redshifted 21-cm fluctuations \citep[e.g.,][]{tozzi2000,furl2004b,loeb2004,iliev2006,wm2007}. In this paper we focus on the statistical signature of reionisation in 21-cm power spectra.

Due to theoretical uncertainties and a lack of observational evidence the exact role quasars play in the reionisation process is uncertain. Although relatively short lived ($t_{\rm q} \sim 10^{6-8}$\,yr) \citep[see, e.g.,][for a review]{martini2004}, it is inferred that luminous quasars radiate close to their Eddington limit \citep[][]{kollmeier2006} such that the ionising photon emission rate of luminous quasars can be $\sim 10^{57}$\,s$^{-1}$. The observed quasar population did not supply sufficent UV photons to reionise the Universe at $z > 6$ \citep[e.g.,][]{fan2001,dijkstra2004b,meiksin2005} even though quasars dominate the production of ionising photons at $z \lsim\,\,3$. However, luminous quasars are known to exist at the edge of the reionisation epoch \citep{fan2006}. Moreover, these high-redshift quasars have a large clustering bias \citep{shen2007}, implying that the effect of their ionising contribution on the statistical signature of the epoch of reionisation may be significant during the latter part of the pre-overlap era. The aim of this paper is to investigate the modification of the 21-cm power spectrum by quasars during the epoch of reionisation. Our results suggest that detailed simulations and reionisation models will need to consider quasar contribution to 21-cm power spectra when interpreting future observations.

This paper is organised as follows: we begin in Section~\ref{Semi-numerical ionisation model} by describing the inclusion of stellar reionisation within a semi-numerical scheme to produce realistic fluctuations in the ionised gas distribution, here we also describe the method by which we populate our simulation field with quasars; in Section~\ref{The quasar contribution to reionisation} we dicuss the choice of parameter space in which we explore the effect of quasars on the ionisation state of the IGM; Section~\ref{Statistical signatures of reionisation} describes the statistical signatures used to characterise reionisation; our results, including their epoch dependence, are presented in Section~\ref{Results}; we discuss our conclusions and their consequences for redshifted 21-cm power spectrum measurements in Section~\ref{Discussion}.

Throughout this paper we adopt the set of cosmological parameters determined by the \textit{Wilkinson Microwave Anisotropy Probe} (\textit{WMAP}) \citep{komatsu2008} for a flat $\Lambda$CDM universe: $\Omega_{\rm m} = 0.27$ (dark matter and baryons); $\Omega_{\Lambda} = 0.73$ (cosmological constant); $\Omega_{\rm b} = 0.046$ (baryons); $h = 0.7$ (Hubble constant); $n_{\rm s} = 1$ (primordial spectrum index); $\sigma_8 = 0.8$ (primordial spectrum normalisation). All distances are in comoving units unless stated otherwise.

% ------------------------------------------------------------------------------------------------------------------------------------
\section{Semi-numerical ionisation model}
\label{Semi-numerical ionisation model}
There are two contributors to the reionisation of the IGM that we consider here: stars and quasars. In this section, we begin by summarising the features of our semi-numerical model for the reionisation of a three-dimensional volume of the IGM by galaxies (Section~\ref{stellarionisation}). Only a brief description is given here--we direct the reader to \cite{geil2008} for further details. We then discuss our method of populating a sample volume with quasars using a Monte Carlo algorithm (Section~\ref{Quasar ionisation}) before describing the three-dimensional realisation of the ionisation state of the IGM including the ionising influence of both stars and quasars (Section~\ref{Inclusion of quasars in the semi-numerical scheme}).

% ----------------------------------------------------------------------
\subsection{Stellar ionisation}
\label{stellarionisation}
We begin by simulating the linear matter overdensity field $\delta(\xvec,z) \equiv \rho_{\rm m}(\xvec,z)/\bar{\rho}_{\rm m}-1$ inside a periodic, comoving, cubic region of volume $V = L^{3}$, by calculating the density contrast in Fourier space $\hat{\delta}(\kvec,z)$ corresponding to a $\Lambda$CDM power spectrum \citep{eh1999} linearly extrapolated to a specified redshift $z$.

The semi-analytic model used to compute the relation between the local dark matter overdensity and the ionisation state of the IGM is based on the model described by \cite{wl2007} and \cite{wm2007}. Any model for the reionisation of the IGM must describe the relation between the emission of ionising photons by stars in galaxies and the ionisation state of the intergalactic gas. This relation is non-trivial as it depends on various internal parameters which may vary with galaxy mass. These parameters include the fraction of gas within galaxies that is converted into stars and accreting black holes, the spectrum of the ionising radiation and the escape fraction of ionising photons from the surrounding interstellar medium as well as the galactic halo and its immediate infall region \citep[see][for a review]{loeb2006}. This relation also depends on intergalactic physics.  In overdense regions of the IGM, galaxies will be over-abundant because small-scale fluctuations need to be of lower amplitude to form a galaxy when embedded in a larger scale overdensity \citep{mo1996}. On the other hand, the increase in the recombination rate in overdense regions counteracts this galaxy bias. The process of reionisation also contains several layers of feedback: radiative feedback heats the IGM and results in the suppression of low-mass galaxy formation which delays the completion of reionisation by lowering the local star formation rate. However, this effect is counteracted in overdense regions by the biased formation of massive galaxies.

The evolution of the ionisation fraction by mass $Q_{\delta_R,R}$ of a particular region of scale $R$ with overdensity $\delta_R$ (at  redshift $z$) may be written as
\begin{eqnarray}
\label{history}
\nonumber
\frac{dQ_{\delta_R,R}}{dt} &=& \frac{N_{\rm ion}}{0.76}\left[Q_{\delta_R,R} \frac{dF_{\rm col}(\delta_R,R,z,M_{\rm ion})}{dt} \right.\\
\nonumber
&+& \left.\left(1-Q_{\delta_R,R}\right)\frac{dF_{\rm col}(\delta_R,R,z,M_{\rm min})}{dt}\right]\\
&-&\alpha_{\rm B}Cn_{\rm H}^0\left[1+\delta_R\frac{D_1(z)}{D_1(z_{\rm obs})}\right] \left(1+z\right)^3Q_{\delta_R,R},
\end{eqnarray}
where $N_{\rm ion}$ is the number of photons entering the IGM per baryon in galaxies, $\alpha_{\rm B}$ is the case-B recombination coefficient and $C$ is the clumping factor (which we assume, for simplicity, to be a constant value of 2). The production rate of ionising photons in neutral regions is assumed to be proportional to the collapsed fraction $F_{\rm col}$ of mass in halos above the minimum threshold mass for star formation ($M_{\rm min}$), whereas the minimum halo mass in ionised regions is limited by the Jeans mass in an ionised IGM ($M_{\rm ion}$). We assume $M_{\rm min}$ corresponds with a virial temperature of $10^4$\,K, representing the hydrogen cooling threshold, and $M_{\rm ion}$ to correspond with a virial temperature of $10^5$\,K, representing the mass below which infall from an ionised IGM is suppressed \citep{dijkstra2004a}. In a region of comoving radius $R$ and mean overdensity $\delta(z)=\delta D_1(z)/D_1(z_{\rm obs})$ (specified at redshift $z$ instead of the usual $z = 0$), we use the extended Press-Schechter model \citep{bond1991} to calculate the relevant collapsed fraction,
\begin{eqnarray}
F_{\rm col}(\delta_R,R,z) = \mbox{erfc}{\left[\frac{\delta_c-\delta_R(z)}{\sqrt{2(\sigma^2_{\rm gal}-\sigma_R^2)}}\right]},
\end{eqnarray}
where $\mbox{erfc}(x)$ is the complimentary error function, $\sigma_R^2$ is the variance of the overdensity field smoothed on a scale $R$ and $\sigma^2_{\rm gal}$ is the variance of the overdensity field smoothed on a scale $R_{\rm gal}$, corresponding to a mass scale of $M_{\rm min}$ or $M_{\rm ion}$ (both evaluated at redshift $z$ rather than at $z = 0$).

Equation~(\ref{history}) may be integrated in time as a function of $\delta_R$.  At a specified redshift this yields the filling fraction of ionised regions within the IGM on various scales $R$ as a function of overdensity. This model predicts the sum of astrophysical effects to be dominated by galaxy bias, and that as a result, overdense regions are reionised first. This leads to the growth of \HII regions via a phase of percolation during which individual \HII regions overlap around clustered sources in overdense regions of the Universe. We may also calculate the corresponding average 21-cm brightness temperature contrast of gas relative to the cosmic microwave background (CMB) \citep[][]{madau1997},
\begin{eqnarray}
\delta T_{\rm b}(\delta_R,R) &=& 26\,(1-Q_{\delta_R,R}) (1+\delta_{\rm b}) \left(1 - \frac{T_{\gamma}}{T_{\rm s}}\right) \left( \frac{\Omega_{\rm b} h^2}{0.022} \right)  \nonumber \\
&\times&\left( \frac{0.15}{\Omega_{\rm m} h^2} \frac{1+z}{10} \right)^{1/2}\,{\rm mK}.
\end{eqnarray}
Here, the spin temperature $T_{\rm s}$, CMB brightness temperature $T_{\gamma}$, baryonic overdensity $\delta_{\rm b}$ (which we assume to be equal to the dark matter overdensity) and ionisation fraction $Q_{\delta_R,R}$ are calculated for a comoving location $\xvec$ at redshift $z$.  We assume $T_{\rm s} \gg T_{\gamma}$ during the epoch of reionisation \citep{ciardi2003,furl2006} and ignore the enhancement of brightness temperature fluctuations due to peculiar velocities in overdense regions \citep{bharadwaj2005,bl2005}. Peculiar velocities were included in the semi-numerical model of \cite{mesinger2007}, who found their effect to be small on scales $\gsim\;10$\,Mpc.

Throughout this paper we consider a model in which the mean IGM is fully reionised by $z = 6$ \citep{fan2006,gnedin2006,white2003}. In this model we assume that star formation proceeds in halos above the hydrogen cooling threshold in neutral regions of the IGM. In ionised regions of the IGM star formation is assumed to be suppressed by radiative feedback. Having determined the value of the ionised fraction $Q_{\delta_R,R}$ as a function of overdensity $\delta_R$ and smoothing scale $R$, we may now construct the ionisation field.  We employ a similar filtering algorithm to that of \cite{mesinger2007} to determine the mass-averaged ionisation state of each voxel within the simulation cube. This is done by repeatedly filtering the linear overdensity field using a real-space spherical `top hat' of radius $R_{\rm f}$, on scales in the range from $L$ to $L/N$ at logarithmic intervals of width $\Delta R_{\rm f}/R_{\rm f} = 0.1$. This procedure is performed in Fourier space for computational efficiency. For all filter scales the ionisation state of each voxel is determined using $Q_{\delta_R,R}$ and deemed to be fully ionised if $Q_{\delta_R,R} \geq 1$. All voxels within a sphere of comoving radius $R$ centred on these positions are flagged and assigned $Q_{\delta_R,R} = 1$, while the remaining non-ionised voxels are assigned an ionised fraction of $Q_{\delta_R,R_{\rm f,min}}$, where $R_{\rm f,min} = L/N$ corresponds to the smallest smoothing scale. A voxel forms part of an \HII region if $Q_{\delta_R,R}>1$ on any scale $R$.

% ----------------------------------------------------------------------
\subsection{Quasar ionisation}
\label{Quasar ionisation}
The ionising emissivity of the known population of quasars diminishes rapidly beyond $z \gsim\;3$. Furthermore, bright quasars are unlikely to contribute signicantly to the ionising background at $z \gsim\;5$ \citep{shapiro1994,haiman2001,wl2003}. More recent results have constrained the relative quasar contribution (at $z \sim 6$) to the ionising background to be $\lsim$ 14 per cent \citep[see, e.g.,][]{srbin2007}. However, while the reionisation of cosmic hydrogen is believed to have been dominated by stars, the exact contribution made by quasars is poorly understood. Moreover, the large clustering bias of high-redshift quasars \citep{shen2007} implies that even a small contribution to reionisation by quasars could have a significant effect on the statistics of 21-cm fluctuations. In this section, we describe the method by which we populate our simulation field with quasars and include their contribution to reionisation in our model.

% ---------------------------------
\subsubsection{Quasar population}
\label{Quasar population}
In this paper, we model the quasar population as follows: all quasars are assumed to reside in dark matter halos and a given halo can host no more than one (active or dormant) quasar at a time; we assume the luminosity of a quasar during its active phase to be related to the mass of its host halo, with a resulting ionisation field that is isotropic. A mass-luminosity relationship is plausible on both theoretical and empirical grounds \citep[see, e.g.,][]{martini2001,kollmeier2006}, however, the model we advocate here is phenomenologically motivated and not intended to encompass the physics of super-massive black hole growth. We do not consider the effects of anisotropic ionising sources.

% ---------------------------------
\subsubsection{Space density of quasars}
\label{Space density of quasars}
If active quasars reside in a fraction $f_{\rm on}$ of halos, then the comoving number density of active quasars $\phi(m,z)$ in host halos with masses between $(m,m + dm)$ at redshift $z$ can be written as $\phi(m,z)dm = f_{\rm on}n_{\rm h}(m,z)dm$, where $n_{\rm h}(m,z)$ is the Sheth-Tormen mass function \citep{sheth2001} at redshift $z$. However, the effect of quasars on the history of reionisation is sensitive to the cumulative effect of the ionising contribution from quasars on the IGM, and so we are more concerned with the probability that a quasar has irradiated its local volume at some time in its past. We therefore write
\begin{eqnarray}
\label{nq2}
\phi(m,z)dm = n_{\rm h}(m,z)dm
\end{eqnarray}
as the comoving number density of quasars (active or dormant) in host halos with masses between $(m,m + dm)$ at redshift $z$. We include fossil quasar-generated \HII regions by assuming they remain ionised after the quasar has become dormant. The validity of this assumption is discussed in Section~\ref{Recombination and fossil HII regions}.
\begin{figure}
\includegraphics[width=8.5cm]{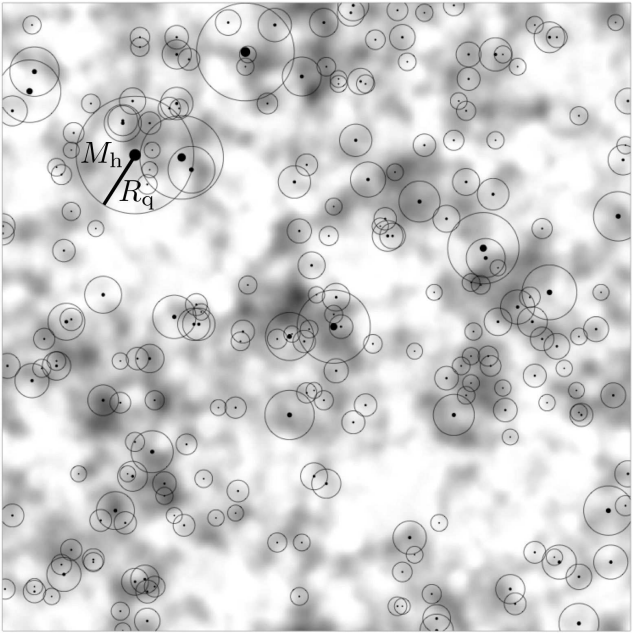}
\caption{Example quasar-populated field showing the target distribution $\pi(X)$ (\textit{greyscale}) and projected quasar positions (\textit{closed circles}) in $100^2$\,Mpc${^2\times 0.4}$\,Mpc simulation slice. \HII regions surrounding quasars are shown as open circles of comoving radius $R_{\rm q}$ corresponding to their extent in a fully neutral IGM. }
\label{QSOfield}
\end{figure}

% ---------------------------------
\subsubsection{Populating the simulation field}
\label{Populating the simulation field}

The strong clustering of galaxies in overdense regions implies that these galaxies, and therefore quasars, should trace the higher density regions of the IGM. To study the impact high-redshift quasars have on the ionisation state of the IGM we populate our simulated field with quasars using a simple case of a random walk Metropolis-Hastings Monte Carlo Markov chain (MCMC) algorithm, which we now describe in its applied context.

The likelihood of a quasar residing at a random location is proportional to the number density of dark matter halos. For small values of large-scale overdensity $\delta$, this number density is proportional to $1+b(M,z)\delta$, where $b(M,z)$ is the galaxy bias.

We denote the Markov chain (MC) by $X = \{X_1,X_2,...,X_{n_{\rm q}}\}$, where $X_n$ is the $n$th quasar placed in the simulation field, of mass $M$ and at position $\xvec_n$. Given the current position of the MC is $X_t$, a move to $Y$ is proposed by randomly perturbing the current position using a spherically symmetric proposal distribution, $q(\xvec,\yvec) = q(|\xvec-\yvec|)$. We accept the proposed move and set $X_t = Y$ based on the acceptance probability
\begin{eqnarray}
\alpha(X_t,Y) = \textrm{min}(1,r_\mathcal M),
\end{eqnarray}
where $r_{\mathcal M}$ is the Metropolis ratio, found by calculating
\begin{eqnarray}
\label{rM}
r_{\mathcal M} \equiv \frac{\pi(Y)q(Y,X_t)}{\pi(X_t)q(X_t,Y)} = \frac{\pi(Y)}{\pi(X_t)}.
\end{eqnarray}
The second equality in Equation~(\ref{rM}) follows from the symmetry of the proposal distribution $q(\xvec,\yvec)$. The target distribution $\pi(X)$ is given by
\begin{eqnarray}
\label{likelihood}
\pi(X) = \frac{1+b(M,z)\delta_R(X)}{1+b(M,z)\delta_{R,\rm max}},
\end{eqnarray}
where $\delta_R$ denotes the overdensity field smoothed on a scale\footnote{As features in the overdensity field of scale$\lsim\,R$ disappear with larger smoothing scales, so do features in the target distribution $\pi(X)$.} $R = 5$\,Mpc. When $b(M,z) = 1$, $\pi(X)$ is simply the smoothed dark matter overdensity at a sample position $\xvec$ normalised with respect to the maximum smoothed dark matter overdensity in the realised volume\footnote{Note that $\pi(X)$ can be negative for $b(M,z)>1$. In such cases we set $\pi(X) = 0$. Furthermore, the normalisation constant for $\pi(X)$ cancels in Equation~(\ref{rM}) and therefore is not strictly required. We include it here to give $\pi(X)$ a probabilistic interpretation (i.e. $0 \leq \pi(X) \leq 1)$.}. We calculate the value for bias using the fitting formula of \cite{sheth2001}.

Populating the density field in the manner described above for a single mass scale generates a sample set (the MC) with a probability density which (in the limit $N_{\rm q}\sim N^3$) converges on the desired target distribution more efficiently than using a transition probability dependent upon $\pi(Y)$ alone. Since the overdensity field (and therefore the target distribution) is multi-modal, an MCMC with a peaked proposal distribution (relative to the spatial scale of the modes in the target distribution) can become bound about a local mode and fail to fully explore other modes of significant probability. We avoid this problem by using a uniform proposal distribution, ensuring that the random walk can reach all potential halo locations. This also removes the necessity of a burn-in period. We execute this algorithm for a range of quasar host halo masses above a minimum $M_{\rm h,min}$. The minimum mass is chosen to be larger than the cosmological Jeans mass in an ionised IGM ($M_{\rm J}$)\footnote{We maintain the same mass bins for all simulations in order to reduce numerical noise in comparisons.} \citep[][]{gnedin2000}.

The large volume simulations required to investigate small frequency Fourier modes (i.e. large spatial scales) in the 21-cm signal at times approaching overlap do not in general resolve this Jeans scale. The simulation has a mass and volume resolution of $M_{\rm cell}$ and $V_{\rm cell}$ respectively, where $M_{\rm cell} = \bar{\rho}_{\rm m}V_{\rm cell}$. A host halo of mass $M_{\rm h} < M_{\rm cell}$ may or may not contain a quasar which is capable of fully ionising the simulation cell. This sub-cell ionisation issue is treated by determining the minimum halo mass required to ionise a simulation cell, $M_{\Delta}$. If $M_{\rm h} > M_{\Delta}$ the cell is fully ionised by the quasar, if $M_{\rm h} < M_{\Delta}$ the cell is only partially ionised and the quasar contribution for that cell (added to the stellar contribution) is $Q_{\rm q} = V_{\rm}/V_{\rm cell}$. The model used to calculate $M_{\Delta}$ is outlined in Section~\ref{Inclusion of quasars in the semi-numerical scheme}. For example, using a $128^3$ grid with a sidelength $L = 300$\,Mpc does not resolve $M_{\rm J}$ at $z = 7$, however, a $5.1\times10^{10}$\,$\Msolar$ halo housing a quasar would fully ionise a volume $V \gsim\;V_{\rm cell}$. The maximum mass scale considered here is $M_{\rm h,max} = 10^{13}$\,$\Msolar$, although the \textit{effective} maximum mass scale is $\sim 2.2\times10^{12}$\,$\Msolar$ (which is the maximum mass of a halo expected to be found in a volume of $300^3$\,Mpc$^3$ at $z \approx 6$).

% ----------------------------------------------------------------------
\subsection{Inclusion of quasars in the semi-numerical scheme}
\label{Inclusion of quasars in the semi-numerical scheme}

In a region of comoving radius $R$ containing a quasar, the cumulative number of ionisations per baryon will be larger than predicted by Equation~(\ref{history}). We incorporate the ionising effect of a quasar by first computing the fraction of the IGM within a region of comoving radius $R$ centred on the comoving position $\xvec$ that has already been reionised by stars [using Equation~(\ref{Semi-numerical ionisation model})], we then add an additional ionisation fraction equal to the quasar's contribution,
\begin{eqnarray}
Q_{\rm q}(\xvec,M) = \bigg[\frac{|\xvec-\xvec_{\rm q}|}{R_{\rm q}(M)}\bigg]^{-3},
\end{eqnarray}
where $R_{\rm q}(M)$ is the comoving radius of the \HII region centred on $\xvec_{\rm q}$ that would have been generated by the quasar alone in a fully neutral and uniform IGM. In contrast to stellar ionisation the quasar contribution comes from a point source, and so the contribution to $Q$ originates from a single voxel only (rather than all voxels within $|\xvec-\xvec_{\rm q}|$). Following this addition we filter the ionisation field as described in Section~\ref{stellarionisation}.

Consider a quasar located within a dark matter halo of mass $M_{\rm h}$, radiating with a line-of-sight ionising photon emission rate of $\dot{N}_{\gamma}$ photons per second. Assuming a uniform IGM and neglecting recombinations, the proper radial extent of the quasars \HII region observed along the line of sight at time \textit{t} is \citep{white2003,haiman2002,fan2006}
\begin{eqnarray}
\label{RqWhite}
R = 3.2\,x_{\HIeqn}^{-1/3}\left(\frac{\dot{N}_{\gamma}}{10^{57}}\frac{t_{\rm age}}{10^7 \textrm{yr}}\right)^{1/3}\left(\frac{1+z}{7.5}\right)^{-1}\textrm{ pMpc},
\end{eqnarray}
where $x_{\HIeqn}$ is the neutral fraction and $t_{\rm age} = t-R(t_{\rm age})/c$ is the age of the quasar corresponding to the time when photons that reach $R$ at time $t$ were emitted. The value of $R$ is subject to large uncertainties, including the quasar lifetime and luminosity, as well as the clumpiness and ionisation state of the surrounding IGM. Because of the method used to include the ionising effect of quasars at locations beyond the sphere that would be generated by the quasar alone in a fully neutral and uniform IGM, we put $x_{\HIeqn} = 1$ when calculating $R$.

We assume that the mass of the central black hole (BH) scales as a power law with the circular velocity $v_c$ of the host halo, $M_{\rm bh} \propto v_c^{\gamma}$ \citep{wl2003}. The circular velocity of a halo of mass $M_{\rm h}$ at redshift $z$ varies as $v_c(z) \propto M_{\rm h}^{1/3}(1+z)^{1/2}$ \citep{bl2001}, which leads to a power law relation between BH mass and halo mass, $M_{\rm bh} \propto M_{\rm h}^{5/3}(1+z)^{5/2}$. If the BH powering this quasar shines at a fraction $\eta$ of its Eddington luminosity then we have $L_{\rm q} \propto \eta M_{\rm bh} \propto \eta M_{\rm h}^{\gamma/3}(1+z)^{\gamma/2}$, where $\gamma = 5$ for simple models of feedback-limited accretion \citep{wl2003}. Since $L_{\rm q} \propto \dot{N_{\gamma}}$, using Equation~(\ref{RqWhite}) leads to (in proper units)
\begin{eqnarray}
\label{2ndlastRq}
R \propto \left[\frac{\eta\,t_{\rm age}(z)}{x_{\HIeqn}}\right]^{1/3}M_{\rm h}^{\gamma/9}(1+z)^{\gamma/6-1}.
\end{eqnarray}
In models where BH growth is limited by feedback over the gas dynamical time \citep[e.g.,][]{wl2003}, the quasar lifetime $t_{\rm q}$ is proportional to $H$, and so $t_{\rm q} \propto (1+z)^{-3/2}$ for $z \gg 1$. Equating $t_{\rm age}$ with $t_{\rm q}$ and substituting into Equation~(\ref{2ndlastRq}) gives
\begin{eqnarray}
\label{Rq}
R_{\rm q}(M_{\rm h},z) = R_0\eta^{1/3}\left(\frac{M_{\rm h}}{10^{12}\,\Msolar}\right)^{5/9}\left(\frac{1+z}{7}\right)^{-2/3},
\end{eqnarray}
Here we adopt a quasar luminosity fraction of $\eta = 0.1$ \citep{martini2001} and an index value $\gamma = 5$ \citep{wp2006}. Since the relative quasar contribution to reionisation is expected to be $\lsim$ 14 per  cent \citep{srbin2007}, we have chosen to investigate fiducial relative quasar ionisation contributions of 10 per cent. We therefore choose $R_0$ and $M_{\rm h,min}$ such that at overlap ($z_{\rm ov} = 6$ in our model) the ionising contribution from a population of quasars alone relative to the ionising contribution of galaxies takes a value $Q_{\rm q}/Q_{\star} = 0.1$.

\begin{figure}
\includegraphics[width= 8.5cm]{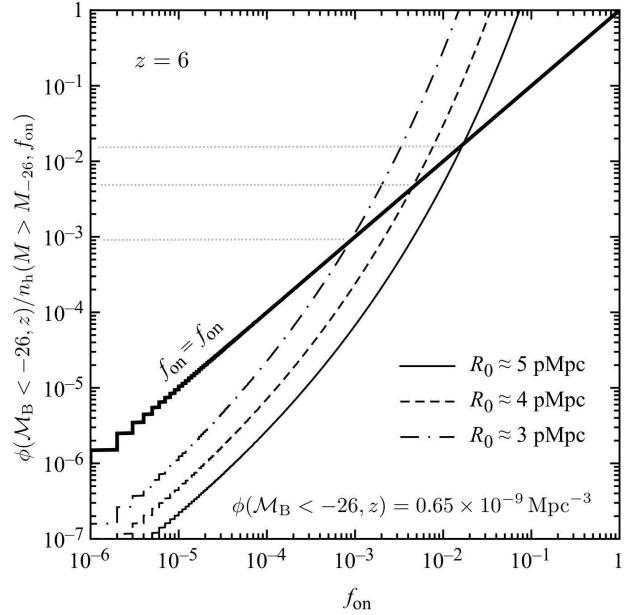} 
\caption{$f_{\rm on}$ parameter solutions for simulations S1, S2 and S3.}
\label{fon}
\end{figure}

We reiterate that the prescription outlined in this section is intended to provide a parameterised model, loosely based on underlying physical models, rather than a complete physical model of all the processes involved. The parameterisation chosen provides a general set of models within which various different possible physical models are contained. The plausibility of this modelling can be investigated by determining the values of various physical parameters for each of our simulation, which we denote by S1, S2 and S3 (in order of decreasing minimum halo mass). We do this for quasar duty cycle $f_{\rm on}$ by noting that the fraction of active quasars with absolute bolometric magnitude $\mathcal{M}_{\rm B} \leq -26$ is
\begin{eqnarray}
\label{eq:fon}
f_{\rm on} = \frac{\psi(\mathcal{M}_{\rm B} \leq -26,z)}{\int_{M_{-26}(f_{\rm on})}^{\infty}dm\frac{\partial n_{\rm h}(m,z)}{\partial m}},
\end{eqnarray}
where $M_{-26}$ is the host halo mass corresponding to a quasar with absolute bolometric magnitude $\mathcal{M}_{\rm B} = -26$. In order to solve for $f_{\rm on}$ we note that for $z \gg 1$,
\begin{eqnarray}
\label{tq}
t_{\rm q} \approx 9.78\times 10^9\,h^{-1}\Omega_{\rm m}^{-1/2}f_{\rm on}(1+z)^{-3/2}\,{\rm yr}.
\end{eqnarray}
Equating $R$ in Equation~(\ref{RqWhite}) with $R_{\rm q}$ in Equation~(\ref{Rq}) enables $M_{\rm h}$ to be written as a function of $f_{\rm on}$ and the model dependent $R_0$, which can then be substituted in Equation~(\ref{eq:fon}). We use $\psi(\mathcal{M}_{\rm B} \leq -26) = 0.65 \times 10^{-9}$\,Mpc$^{-3}$ at $z = 6$ \citep{fan2001} and $\dot{N}_{\gamma}(\mathcal{M}_{\rm B} = -26) = 2 \times 10^{57}$\,s$^{-1}$ \citep{white2003}. Figure~\ref{fon} shows the solution for $f_{\rm on}$ from Equation~(\ref{eq:fon}) graphically for $R_0 \approx 3$, 4 and 5\, proper Mpc. Table~\ref{tab:param} gives the values of these parameters used in simulations S1--S3. We find duty cycles of $\sim 10^{-3}$--$4\times 10^{-2}$, which at $z \approx 6$ corresponds to $t_{\rm q} \sim 10^6$--$4\times 10^7$\,yr. These are comparable to the Salpeter time for doubling of BH mass and to estimates of quasar lifetime \citep[][]{martini2004}.

% ---------------------------------
\subsubsection{Recombination and fossil \HII regions}
\label{Recombination and fossil HII regions}

The prescription outlined above does not consider the effect of recombinations within quasar-generated \HII regions. By investigating recombinations in inhomogeneous fossil ionised regions around quasars, \cite{furl2008} established that most of the \HII regions formed during hydrogen reionisation remain highly ionised throughout the entire reionisation process. They found that the ionising background from galaxies inside the fossil \HII regions, together with any residual low-level emission from the BH past its bright quasar phase, efficiently suppresses recombinations during hydrogen reionisation. Thus, even though quasars are thought to be transient, the assumption of fully ionised relic bubbles yields a good approximation at the level of our model.

% ------------------------------------------------------------------------------------------------------------------------------------
\section{The quasar contribution to reionisation}
\label{The quasar contribution to reionisation}

\begin{figure}
\includegraphics[width= 8.5cm]{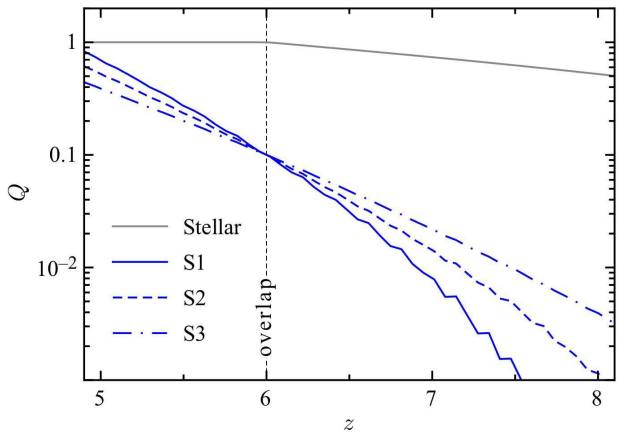}
\caption{Evolution of the upper-limit global ionised fraction $Q$ due to the ionising contribution from the stellar component of galaxies only and quasars only. Quasar-only contributions are shown using $\gamma = 5/3$ and a Sheth-Tormen mass function. The vertical grey line shows the redshift of overlap for this model, $z_{\rm ov} = 6$.}
\label{Qevolution}
\end{figure}

We investigate the ionising contribution of quasars and the 21-cm power spectrum as a function of redshift based on the parameterised model in Equation~(\ref{Rq}), together with the halo mass function and a minimum halo mass $M_{\rm h,min}$. Table~\ref{tab:param} gives the parameter values used in our simulations, which correspond to the plots in Figure~\ref{Qevolution} which shows the resulting contributions of stars and quasars to reionisation. Figure~\ref{Qevolution} shows that the assumption of a lower minimum halo mass allows quasars to contribute more significantly at earlier phases of reionisation.

\begin{figure*}
\includegraphics[width= 18cm]{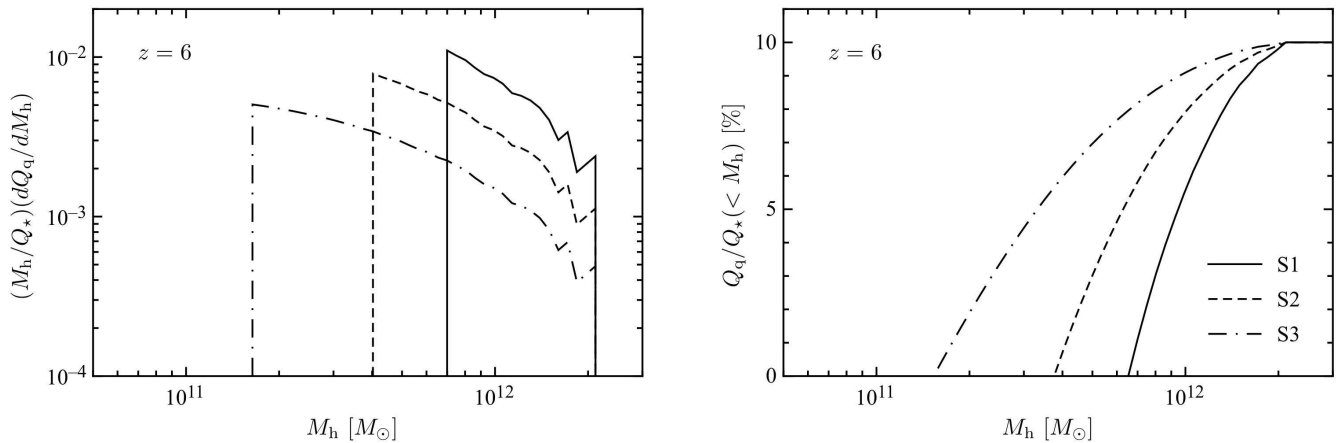}
\caption{Differential (\textit{left}) and cumulative (\textit{right}) ionisation fraction due to quasars $Q_{\rm q}$ as a function of $M_{\rm h}$ for the simulation models considered.}
\label{QvsM}
\end{figure*}

Figure~\ref{QvsM} shows the relative differential ionised fraction due to quasars $(M_{\rm h}/Q_{\star})dQ_{\rm q}/dM_{\rm h}$ and the cumulative ionised fraction due to quasars $Q_{\rm q}/Q_{\star}(<M_{\rm h})$ as a function of $M_{\rm h}$ for simulations S1--S3. The differential plots demonstrate that the bulk of the ionisation by quasars is achieved by quasars in lower-mass host halos, despite larger bubbles being generated by quasars in halos of higher mass. This is due to the greater population of lower-mass halos. As mentioned above, the parameter sets used for these simulations have been chosen so as to produce a cumulative contribution to reionisation by quasars of 10 per cent. We also carried out simulations with a cumulative relative ionisation contribution by quasars of $\approx$ 3 per cent. The results of these simulations (not presented) demonstrate that the effect on the resulting power spectra scale in proportion to quasar contribution.

\begin{table*}
\centering
\caption{Parameter values used in our simulations. Simulation labels refer to the panel and numbered datam in Figure~\ref{Qevolution}. $M_{X}$ denotes a halo mass in units of $10^{X}$\,$\Msolar$. ($N = 128$, $L = 300$\,Mpc)}
\begin{tabular}{c c r c c c c c c c}
\hline\hline
Sim$^{\rm \underline{n}}$& $Q_{\rm q}/Q_{\star}|_{z_{\rm ov}}$  & $M_{\rm h,min}$ & $R_0$ (pMpc) & $f_{\rm on}$ & $t_{\rm q}$ ($10^7$\,yr) & $N_{\rm q}(z=7)$ & $N_{\rm q}(z=8)$ & $N_{\rm q}(z=9)$ & $N_{\rm q}(z=10)$\\
\hline
S1 & 10\% & 6.6\,$M_{11}$ & 5.1 & 0.04 & 5.8 & 14 & 0 & 0 & 0\\
S2 & 10\% & 3.8\,$M_{11}$ & 3.9 & 0.005 & 0.73 & 108 & 10 & 0 & 0\\
S3 & 10\% & 1.6\,$M_{10}$ & 3.0 & 0.001 & 0.15 & 1294 & 252 & 34 & 2\\
\hline
\end{tabular}
\label{tab:param}
\end{table*}

% ------------------------------------------------------------------------------------------------------------------------------------
\section{Statistical signatures of reionisation}
\label{Statistical signatures of reionisation}
As mentioned in Section~\ref{Introduction}, statistical observations of the epoch of reionisation promise to provide a wealth of information about the properties of neutral hydrogen at high redshift as well as some of the fundamental astrophysics behind the reionisation process and the first luminous objects. While density perturbations in the matter distribution mediate fluctuations in the 21-cm signal both prior to and following reionisation, during the reionisation era the relation between the 21-cm power spectrum and the underlying matter power spectrum is complex and, in its late stages, is dominated by the formation of large ionised ``bubbles" \citep{furl2004b,mcquinn2006}. These bubbles of ionised hydrogen imprint features on the 21-cm power spectrum that reflect the luminosity and clustering of ionising sources responsible for reionisation. The 21-cm power spectrum can be used in a statistical test to distinguish candidate reionisation models, as well as to constrain the history and morphology of reionisation \citep{barkana2008}.

In this paper, we employ the dimensionless power spectrum $\Delta_{21}^2(k,z)$ as the key statistical measure, which is the contribution to the variance of the redshifted 21-cm brightness temperature contrast $\delta T_{\rm b}(z)$ per logarithmic interval in wave-number $k = 2\pi/\lambda$. This measure is related to the dimensional form of the power spectrum $P_{21}(k,z)$ by
\begin{eqnarray}
\label{Deltasq}
\Delta_{21}^2(k,z) = \frac{1}{(2\pi)^3} 4\pi k^3P_{21}(k,z).
\end{eqnarray}
$P_{21}(k,z)$ is estimated by averaging over all $m$ modes of the Fourier transform ($\hat{T}$) of $T(z)$ in a thin spherical shell in $k$-space,
\begin{eqnarray}
P_{21}(k,z) \equiv \langle|\hat{T}(\kvec,z)|^2\rangle_k = \frac{1}{m}\sum _{i=1}^m |\hat{T}_i(k,z)|^2.
\end{eqnarray}
Note that we must use caution when employing this statistical measure since the spherical symmetry of the signal is broken by redshift evolution over certain ranges of redshift. This issue has been discussed by a number of authors \citep[see, e.g.,][]{morales2004,bl2005,mcquinn2006} and its effect on the sensitivity of the 21-cm power spectrum has been investigated by \cite{mcquinn2006}. Our numerical power spectrum measurements are subject to sample variance, arising from the finite number of independent modes counted in each $k$-shell, which corresponds to the finite number of independent wavelengths $\lambda = 2\pi/k$ that can fit into the simulated volume\footnote{In general, a measured power spectrum will also have a component of Poisson noise. Poisson noise is a discreteness effect that is present due to imperfect sampling of the field, for example the finite number of galaxies in survey, or the gridding procedure used in creating inital conditions for $N$-body simulations. Poisson noise is not present in our power spectra since we create our mass distributions using a fluid approximation. Rather than randomly place discrete massive particles in the simulation box in accordance with the required power spectrum of density fluctuations.}.

% ------------------------------------------------------------------------------------------------------------------------------------
\section{Results}
\label{Results}
\begin{figure*}
\includegraphics[width= 18.0cm]{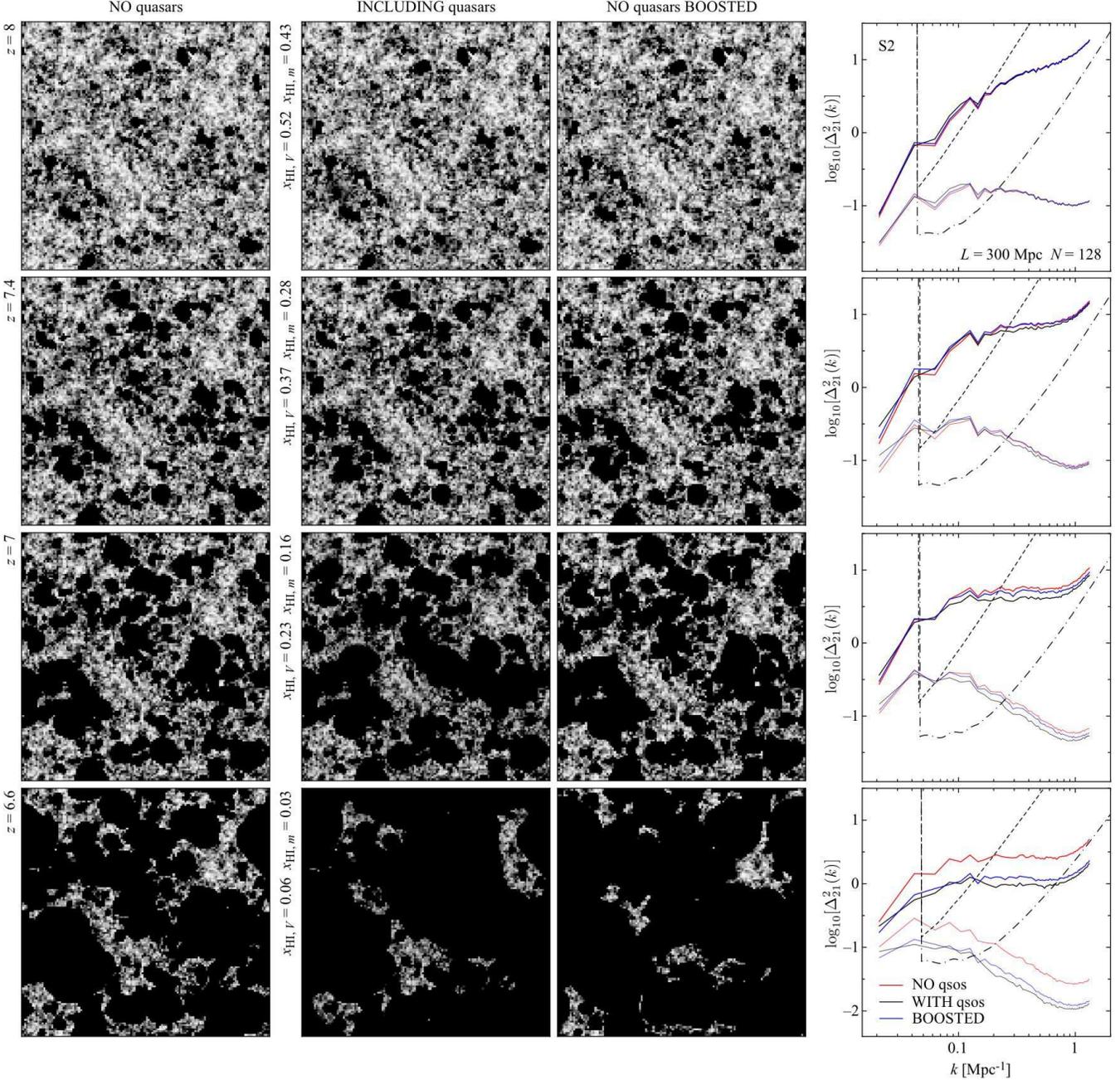}
\caption{Multi-phase ionisation maps for simulation S2 (see Table\,\ref{tab:param}) at $z = 6.6$, 7, 7.4 and 8 (\textit{bottom to top}) with (\textit{second column}) and without (\textit{first column}) the ionising influence of quasars, and with boosted ionisation fraction at $x_{\HIeqn} = 0.03$, 0.16, 0.28 and 0.43 (\textit{bottom to top}). Each slice has a side length of 300\,Mpc and is $\approx 2.3$\,Mpc deep. \textit{Fourth column:} Spherically averaged three-dimensional 21-cm brightness temperature power spectra with (\textit{black}) and without (\textit{red}) quasars, and boosted (\textit{blue}). Corresponding errors are shown below the spectra (\textit{faint}). Fiducial sensitivity curves are shown for 1000\,hr integrations using the MWA (\textit{dash}) and the MWA5000 (\textit{dot-dash}) with four 8\,MHz processed sub-bands.}
\label{128results}
\end{figure*}

We begin our analysis by presenting one example of the evolution of the IGM during the reionisation era with and without the ionising effect of quasars. The left-hand and central left-hand panels of Figure~\ref{128results} show multi-phase ionisation maps for simulation S2 at $z = 6.6$, 7, 7.4 and 8, with and without the ionising influence of quasars respectively. The right-hand panel shows the corresponding power spectra and sample variance (faint lower curves). It is evident, both in the slices shown as well as in the corresponding averaged global ionisation fraction, that the simulated volumes with quasars are more ionised than those without at the same redshift. The resulting impact of the quasar-generated contribution to ionisation on the 21-cm power spectra is to decrease the overall power on all scales. The effect is epoch dependent since the accumulated quasar contribution to reionisation is redshift dependent (as shown in Figure~\ref{Qevolution}). The modification of the spectra arises due to the redistribution of fluctuation power between wave-modes, originating from the spatial deformation of the stellar-generated bubbles by the quasar emission. In addition to the modification of power, the addition of quasar ionisations also modifies the slope of the power spectra. This effect is most easily seen by calculating the relative amplitude between the power spectra without and with quasars,
\begin{eqnarray}
\label{Deltasqrelkz}
\Delta^2_{\rm{rel}}(k,z) = \frac{\Delta^2_{21,\rm{no\,QSO}}(k,z)}{\Delta^2_{21,\rm{QSO}}(k,z)}.
\end{eqnarray}

\begin{figure*}
\includegraphics[width= 16cm]{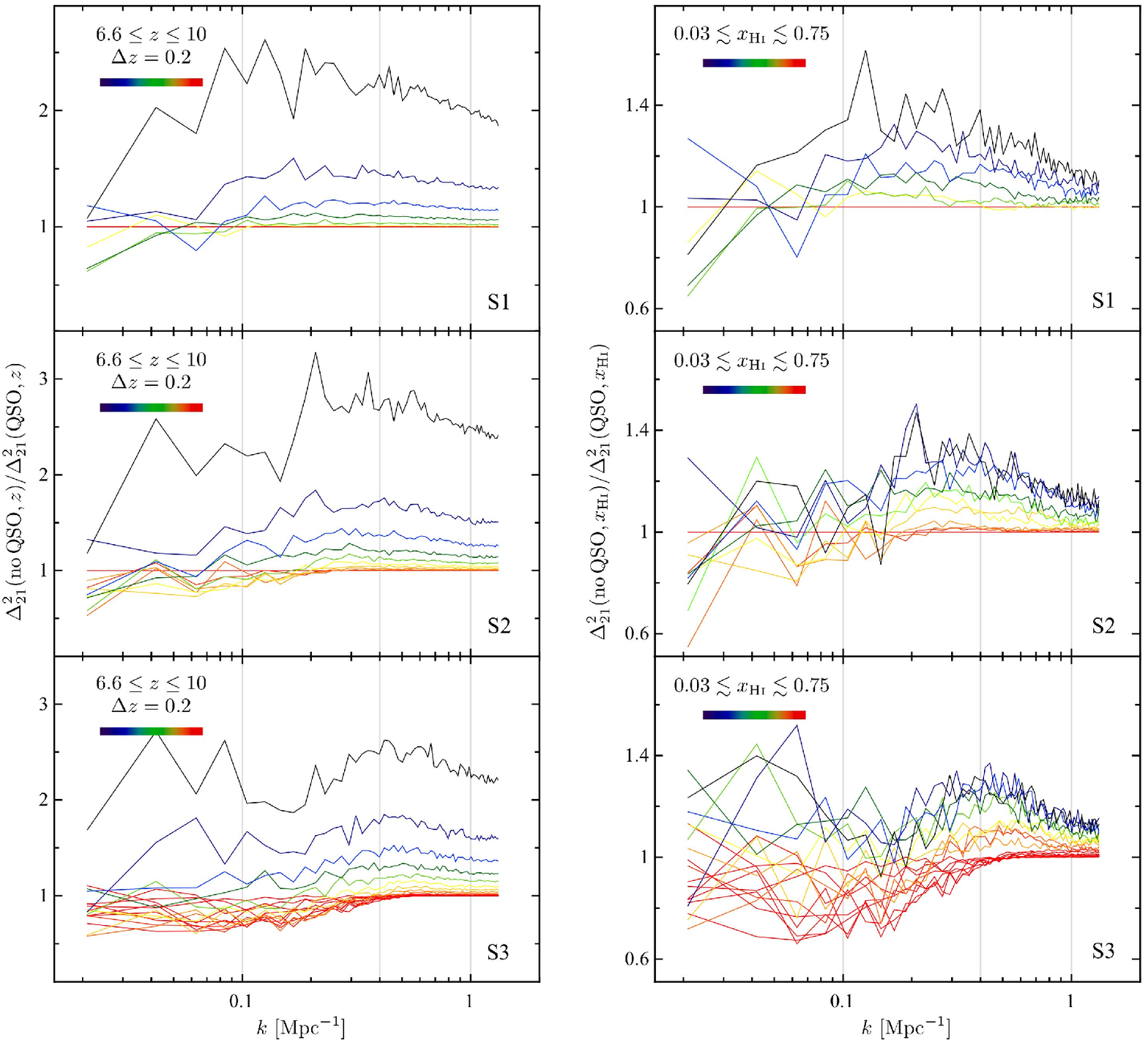}
\caption{Comparisons between power spectra for simulations S1 (\textit{top}), S2 (\textit{middle}) and S3 (\textit{bottom}) with and without the inclusion of effects by quasars for $6.6 \leq z \leq 10$ (\textit{left}) and $0.03 \lsim\,\,x_{\HIeqn} \lsim\,\,0.73$ (\textit{right}). The horizontal line marks unity which indicates equal power for both scenarios. The vertical lines corresponds to $k \approx 0.1$, 0.4, 1\,Mpc$^{-1}$, or a spatial scale of $\approx 60$, 15, 6\,Mpc respectively.}
\label{comp_comp_10S_z_xHI_128}
\end{figure*}

The resulting relative power for simulations S1, S2 and S3 are shown in the left-hand panels of Figures~\ref{comp_comp_10S_z_xHI_128} and \ref{comp_comp_k}. Figure~\ref{comp_comp_10S_z_xHI_128} shows the relative power at different redshifts as a function of wave-number. Figure~\ref{comp_comp_k} shows cuts of the data as a function of redshift at three different scales corresponding to wave-numbers $k \approx 0.1$, 0.4 and 1\,Mpc. The dependence on scale is discussed further in Section~\ref{Discussion}.

% ----------------------------------------------------------------------
\subsection{The effect of quasars at constant neutral fraction}
\label{Neutral fraction dependence}
In the examples above, the addition of quasar ionisation lowers the neutral fraction at a fixed redshift, and hence the comparison in Equation~(\ref{Deltasqrelkz}) is not made at an equivalent stage of reionisation. Therefore, we next isolate the effect of quasars as a function of neutral fraction by removing the difference in overall power. This is achieved by boosting the stellar ionising efficiency in the stellar-only simulations so as to match the mass-weighted global neutral fraction of the simulation including quasars. The resulting ``boosted'' realisations for simulation S2 (to match $x_{\HIeqn} = 0.03$, 0.16, 0.28 and 0.43 in the examples) are shown in the central right-hand panels of Figure~\ref{128results}, and their corresponding power spectra and sample variance in the right-hand panels.

The effect of quasars on the simulated 21-cm power spectra as a function of neutral fraction (and $M_{\rm h,min}$) is again determined by calculating the relative power,
\begin{eqnarray}
\Delta^2_{\rm{rel}}(k,x_{\HIeqn}) = \frac{\Delta^2_{21,\rm{no\,QSO}}(k,x_{\HIeqn})}{\Delta^2_{21,\rm{QSO}}(k,x_{\HIeqn})},
\end{eqnarray}
where, as mentioned, $\Delta^2_{21,\rm{no\,QSO}}$ has been calculated using the boosted simulations so as to match the mass-weighted neutral fraction of the simulations including quasars.

\begin{figure*}
\includegraphics[width= 18cm]{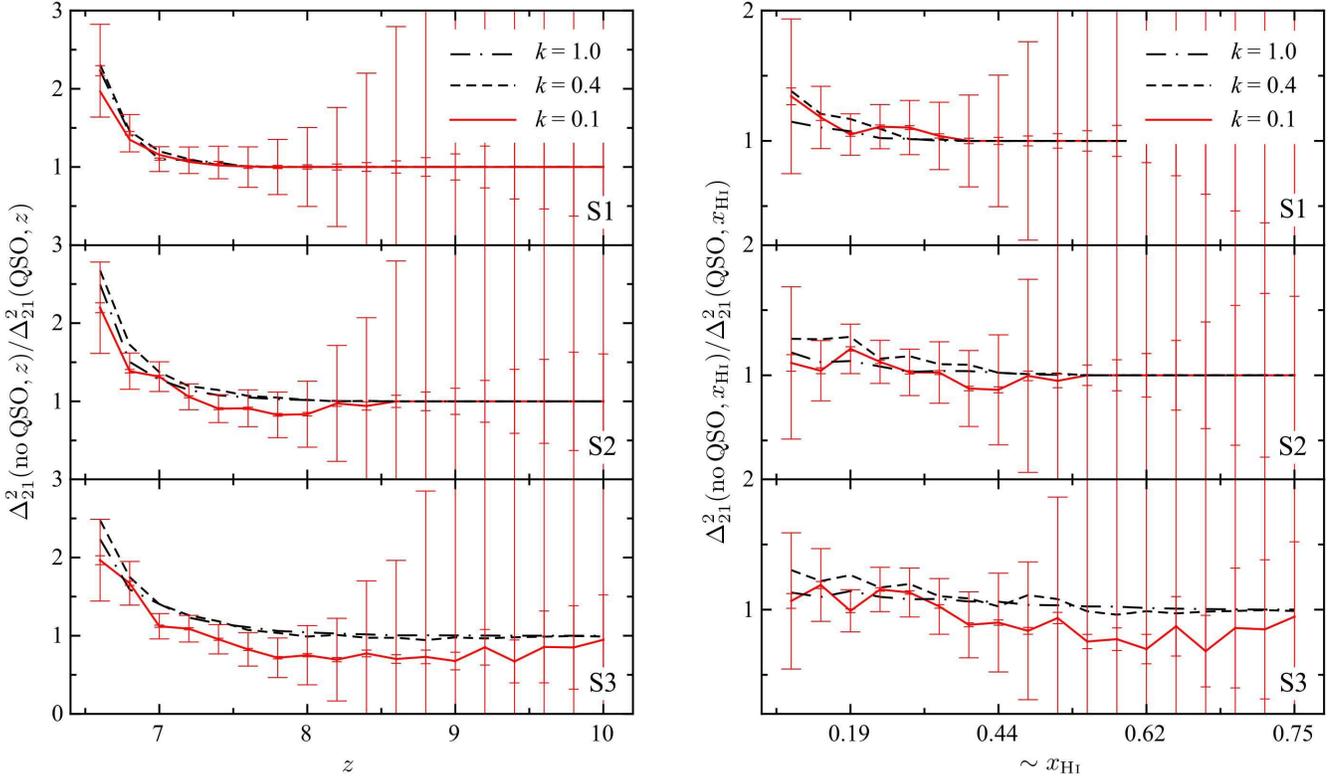}
\caption{Comparisons between 21-cm power spectra with and without including the ionising of effect of quasars as a function of redshift (\textit{left}) and neutral fraction (\textit{right}) for simulations with $Q_{\rm q}/Q_{\star}|_{z_{\rm ov}} = 0.1$. Three plots are shown for each simulation: $k \approx 1$ (\textit{dot-dashed}), $k \approx 0.4$ (\textit{dashed}) and $k \approx 0.1$\,Mpc$^{-1}$ (\textit{solid red}). Instrumental sensitivity levels of the MWA (\textit{wide bar}) and MWA5000 (\textit{narrow bar}) for 1000\,hr integrations are shown for $k \approx 0.1$\,Mpc$^{-1}$ only.}
\label{comp_comp_k}
\end{figure*}

The resulting relative power for simulations S1, S2 and S3 are shown in the right-hand panels of Figures~\ref{comp_comp_10S_z_xHI_128} and \ref{comp_comp_k}. Figure~\ref{comp_comp_10S_z_xHI_128} shows the relative power at different neutral fractions as a function of wave-number. Figure~\ref{comp_comp_k} shows cuts of the data as a function of neutral fraction at three different scales corresponding to wave-numbers $k \approx 0.1$, 0.4 and 1\,Mpc. The effect of the quasars is scale-dependent, indicating that the additional bias of the quasars makes fewer larger bubbles than a corresponding increase in stellar flux.

% ------------------------------------------------------------------------------------------------------------------------------------
\section{Discussion}
\label{Discussion}

Our results indicate that a 10 per cent contribution to reionisation from quasars leads to a damping of the 21-cm power spectrum by a factor of up to 2 at constant redshift, and by up to 30 per cent at constant neutral fraction shortly before the end of reionisation ($\Delta z \sim 1$). The level of damping is both a function of scale and redshift.

Firstly, in all simulations and at all scales, the effect on the relative power is more prominent at later epochs when there is a larger quasar population. At these later epochs, the cumulative quasar contribution to reionisation (from both newly formed and fossil \HII regions) is greater (see Figure~\ref{Qevolution}). Moreover, the rate at which the modification evolves is more rapid in models with more massive host halos, resulting from rapid evolution of the halo mass function.

The scale dependence of the relative power can be attributed to the effects of halo bias. As discussed, our results show that quasars increase the ionisation fraction at fixed redshift. This leads to a decrease in the amplitude of the power spectrum late in the reionisation era, since the growth of \HII regions removes signal at a rate that counteracts the increase in power resulting from the \HII region-induced structure in the IGM. This is the ``fall" of 21-cm fluctuations described by \cite{lidz2008}. The effect is larger on small scales, since the addition of quasars increases the size of the \HII regions, owing to quasars being biased to overdense regions which reionise first. This removes more of the sources of small-scale power. The effect is less significant on large scales, where removal of the source of large-scale power due to the ionisation of gas is partly counteracted by its increase from larger \HII regions.

In the case of comparison at constant neutral fraction the opposite is true. By increasing the contribution of star formation to match the neutral fraction, the fraction of small-scale power removed by ionisation is the same for stars and quasars. As a result $\Delta^2_{\rm{rel}} \sim 1$ for large $k$. However, quasars result in fewer larger ionised regions as they have a higher bias than the galaxies. Late in reionisation this accelerates the decline in power (during the ``fall'') and results in $\Delta^2_{\rm{rel}} > 1$ for intermediate values of $k$. Thus, the level of modification of the shape of the 21-cm power spectrum can be related to the host halo masses of the quasars.

Any radio interferometer is subject to instrumental noise and has limited sensitivity arising from the finite volume of the observation. We consider observational parameters corresponding to the design specications of the MWA, and of a hypothetical follow-up to the MWA (which we term the MWA5000). We assume a continuous, circularly symmetric distribution of 512 antenna tiles for the MWA. This distribution has a constant antenna density core of radius $r_{\rm c} \approx 28$\,m and an inverse square radial antenna density profile for $r > r_{\rm c}$ (giving a maximum radius of 0.75\,km). Each antenna tile contains 16 cross-dipoles to yield an effective collecting area of $A_{\rm e} = 16(\lambda^2/4)$ (the area is capped for $\lambda > 2.1$\,m). The MWA 5000 is assumed to follow the basic design of the MWA. The quantitative differences are that we assume the MWA5000 to have 5000 tiles within a diameter of 2\,km, with a flat antenna density core of radius 40\,m. In each case, we assume one field is observed for an integrated time of 1000\,hr. Following the work of McQuinn et al. (2006) we assume that foregrounds can be removed over 8\,MHz sub-bands, within a bandpass of 32\,MHz [foreground removal therefore imposes a minimum accessible wave-number of $k_{\rm min} \approx 0.04[(1+z)/7.5]^{-1}$\,Mpc$^{-1}$].

We compute the sensitivity with which the effect of quasars on the shape of the 21-cm power spectum could be detected following the procedure outlined by \cite{mcquinn2006} and \cite{bowman2006} \citep[see also ][]{wlg2008}. Written in terms of the cosmic wave-vector $\kvec = \kvec_\parallel + \kvec_\perp$, the resulting error in the 21-cm power spectrum \textit{per mode} is
\begin{eqnarray}
\delta P_{21}(\kvec) = \frac{T_{\rm sys}^2}{Bt_0} \frac{D^2 \Delta D}{n_{\rm b}(U,\nu)} \left(\frac{\lambda^2}{A_{\rm e}}\right)^2  + P_{21}(\kvec),
\end{eqnarray}
where $T_{\rm sys} \approx 250[(1+z)/7]^{2.6}$\,K is the system temperature of the instrument, $D(z)$ is the comoving distance to the point of emission at redshift $z$, $\Delta D$ is the comoving depth of the survey volume corresponding to the bandwidth $B$, $t_0$ is the total integration time, $n_{\rm b}(U,\nu)$ is the number density of baselines that can observe the visibility $\Uvec$, where $U = k_\perp D/2\pi$ and $\lambda$ is the observed wavelength.

Although the observed 21-cm power spectrum is not spherically symmetric, it is symmetric about the line of sight. This makes it possible to calculate the overall power spectral sensitivity of the radio interferometer using the Fourier modes contained within an infinitesimal annulus around the line of sight of constant $(k,\theta)$, where $\cos(\theta) = \kvec \cdot \hat{\zvec}/k$ ($\hat{\zvec}$ is  the unit vector pointing in the direction of the line of sight). The power spectral sensitivity over such an annulus is given by
\begin{eqnarray}
\sigma_P(k,\theta) = \frac{\delta P_{21}(k,\theta)}{\sqrt{N_{\rm m}(k,\theta)}},
\end{eqnarray}
where $N_{\rm m}(k,\theta)$ denotes the number of observable modes in the annulus (only modes whose line-of-sight components fit within the observed bandpass are included). In terms of the $k$-vector components $k$ and $\theta$, the number of independent Fourier modes within an annulus of radial width $dk$ and angular width $d\theta$ is $N_{\rm m}(k,\theta) = 2\pi k^2 V \sin(\theta) dk\,d\theta/(2\pi)^3$, where $V = D^2\Delta D(\lambda^2/A_{\rm e})$ is the observed volume. Averaging $\sigma_P(k,\theta)$ over $\theta$ gives the spherically averaged sensitivity to the 21-cm power spectrum $\sigma_P(k)$,
\begin{eqnarray}
\frac{1}{\sigma^2_P(k)} = \sum_{\theta}\frac{1}{\sigma^2_P(k,\theta)}.
\end{eqnarray}
In order to find the sensitivity in terms of $\Delta^2_{21}$ we use Equation~(\ref{Deltasq}), which gives
\begin{eqnarray}
\sigma_{\Delta^2}(k,z) = \frac{1}{(2\pi)^3}4\pi k^3 \left[\sum_{\theta}\frac{1}{\sigma^2_P(k,\theta,z)}\right]^{-1/2}.
\end{eqnarray}
Figure~\ref{128results} shows the fiducial spherically averaged sensitivity curves for the MWA and MWA5000 (within bins of $\Delta k = k/10$) for integration times of 1000\,hr using four 8\,MHz processed sub-bands. The corresponding sensitivities are also included in the relative power plots shown in Figure~\ref{comp_comp_k} for $k \approx 0.1$\,Mpc$^{-1}$ only, where the instrumental sensitivities are near their maximum.

Our results suggest that (for $k \approx 0.1$\,Mpc$^{-1}$), using a constant redshift comparison, the effect of quasars on the 21-cm power spectrum is comparable to the precision achievable by the MWA and the MWA5000 at $z \lsim\,\,7$ for all three simulations, and at $7.5 \lsim\,\,z \lsim\,\,8$ in simulation S3. The sensitivity of the MWA5000 is comparable to the effect up to a higher redshift of $z \approx 9.4$ in simulation S3. When making a comparison at constant neutral fraction, the effect is comparable to the observational precision of the MWA5000 only at $x_{\HIeqn} \lsim\,\,0.25$ (power suppression) in all simulations, and $0.4 \lsim\,\,x_{\HIeqn} \lsim\,\,0.65$ (power amplification) in simulation S3. Thus, with respect to modelling the observable 21-cm power spectrum close to the end of reionisation, our results indicate that the potential contribution to ionisation by quasars will need to be considered.

% ------------------------------------------------------------------------------------------------------------------------------------
\section{Summary}
\label{Summary and conclusions}
In this paper, we have assessed the effect of high-redshift quasars on the 21-cm power spectrum during the epoch of reionisation. Our approach has been to implement a semi-numerical scheme to calculate the three-dimensional structure of ionised regions surrounding massive halos at high redshift. We have included the ionising influence of luminous quasars by populating a simulated overdensity field with quasars using a Monte Carlo Markov Chain algorithm. Different parameterisations of the quasar luminosity-halo mass relationship have been used to analyse the relative effect on spherically averaged power in 21-cm emission between simulations with and without the ionising influence of quasars. This comparison was carried out for simulations at the same redshift, ranging between $6.6 \leq z \leq 10$, as well as at the same global neutral fraction, ranging between $0.03 \lsim\,\,x_{\HIeqn} \lsim\,\,0.75$. Our results show that the cumulative ionising effect of quasars can suppress the 21-cm power spectrum by a factor of up to 2 at constant redshift, and by up to 30 per cent at constant neutral fraction shortly before the end ($\Delta z \sim 1$) of reionisation, both as a function of scale and redshift.

The possible contribution of quasars to reionisation will complicate efforts to interpret observed 21-cm power spectra. For example, quasar ionisation results in modification of the amplitude of the power spectra at fixed neutral fraction. If quasars are ignored, this will result in biased inferences regarding galaxy formation \cite[e.g., using a method as described in][]{barkana2008}. Similarly, the modification of the slope will lead to incorrect inferences regarding bubble size and host mass based on power spectral analyses that do not allow for quasars. Thus, modelling of the 21-cm power spectrum is likely to be degenerate with the quasar population, and so will require input from observations of the high-redshift luminosity functions of stars and/or quasars.\vspace{3mm}

\noindent{\bf Acknowledgments} PMG acknowledges the support of an Australian Postgraduate Award and the hospitality of the Institute for Theory and Computation at the Harvard-Smithsonian Center for Astrophysics, where part of this research was done. The research was supported by the Australian Research Council (JSBW).

\newcommand{\noopsort}[1]{}

\bibliography{bib.bib}

\begin{thebibliography}{}

\bibitem[\protect\citeauthoryear{{Barkana}}{{Barkana}}{2008}]{barkana2008}
{Barkana} R.,  2008, ArXiv e-prints, 0806.2333

\bibitem[\protect\citeauthoryear{{Barkana} \& {Loeb}}{{Barkana} \&
  {Loeb}}{2001}]{bl2001}
{Barkana} R.,  {Loeb} A.,  2001, Phys. Rep., 349, 125

\bibitem[\protect\citeauthoryear{{Barkana} \& {Loeb}}{{Barkana} \&
  {Loeb}}{2005}]{bl2005}
{Barkana} R.,  {Loeb} A.,  2005, ApJL, 624, L65

\bibitem[\protect\citeauthoryear{{Bharadwaj} \& {Ali}}{{Bharadwaj} \&
  {Ali}}{2005}]{bharadwaj2005}
{Bharadwaj} S.,  {Ali} S.~S.,  2005, MNRAS, 356, 1519

\bibitem[\protect\citeauthoryear{{Bond}, {Cole}, {Efstathiou} \&
  {Kaiser}}{{Bond} et~al.}{1991}]{bond1991}
{Bond} J.~R.,  {Cole} S.,  {Efstathiou} G.,    {Kaiser} N.,  1991, ApJ, 379,
  440

\bibitem[\protect\citeauthoryear{{Bowman}, {Morales} \& {Hewitt}}{{Bowman}
  et~al.}{2006}]{bowman2006}
{Bowman} J.~D.,  {Morales} M.~F.,    {Hewitt} J.~N.,  2006, ApJ, 638, 20

\bibitem[\protect\citeauthoryear{{Ciardi} \& {Madau}}{{Ciardi} \&
  {Madau}}{2003}]{ciardi2003}
{Ciardi} B.,  {Madau} P.,  2003, ApJ, 596, 1

\bibitem[\protect\citeauthoryear{{Dijkstra}, {Haiman} \& {Loeb}}{{Dijkstra}
  et~al.}{2004}]{dijkstra2004b}
{Dijkstra} M.,  {Haiman} Z.,    {Loeb} A.,  2004, ApJ, 613, 646

\bibitem[\protect\citeauthoryear{{Dijkstra}, {Haiman}, {Rees} \&
  {Weinberg}}{{Dijkstra} et~al.}{2004}]{dijkstra2004a}
{Dijkstra} M.,  {Haiman} Z.,  {Rees} M.~J.,    {Weinberg} D.~H.,  2004, ApJ,
  601, 666

\bibitem[\protect\citeauthoryear{{Eisenstein} \& {Hu}}{{Eisenstein} \&
  {Hu}}{1999}]{eh1999}
{Eisenstein} D.~J.,  {Hu} W.,  1999, ApJ, 511, 5

\bibitem[\protect\citeauthoryear{{Fan}, {Strauss}, {Becker}, {White}, {Gunn},
  {Knapp}, {Richards}, {Schneider}, {Brinkmann} \& {Fukugita}}{{Fan}
  et~al.}{2006}]{fan2006}
{Fan} X.,  {Strauss} M.~A.,  {Becker} R.~H.,  {White} R.~L.,  {Gunn} J.~E.,
  {Knapp} G.~R.,  {Richards} G.~T.,  {Schneider} D.~P.,  {Brinkmann} J.,
  {Fukugita} M.,  2006, AJ, 132, 117

\bibitem[\protect\citeauthoryear{{Fan et al.}}{{Fan et al.}}{2001}]{fan2001}
{Fan et al.} 2001, AJ, 122, 2833

\bibitem[\protect\citeauthoryear{{Furlanetto}, {Haiman} \& {Oh}}{{Furlanetto}
  et~al.}{2008}]{furl2008}
{Furlanetto} S.,  {Haiman} Z.,    {Oh} S.~P.,  2008, ArXiv e-prints, 0803.3454

\bibitem[\protect\citeauthoryear{{Furlanetto}}{{Furlanetto}}{2006}]{furl2006a}
{Furlanetto} S.~R.,  2006, MNRAS, 371, 867

\bibitem[\protect\citeauthoryear{{Furlanetto}, {McQuinn} \&
  {Hernquist}}{{Furlanetto} et~al.}{2006}]{furl2006}
{Furlanetto} S.~R.,  {McQuinn} M.,    {Hernquist} L.,  2006, MNRAS, 365, 115

\bibitem[\protect\citeauthoryear{{Furlanetto}, {Oh} \& {Briggs}}{{Furlanetto}
  et~al.}{2006}]{furl2006b}
{Furlanetto} S.~R.,  {Oh} S.~P.,    {Briggs} F.~H.,  2006, Phys. Rep., 433, 181

\bibitem[\protect\citeauthoryear{{Furlanetto}, {Zaldarriaga} \&
  {Hernquist}}{{Furlanetto} et~al.}{2004}]{furl2004b}
{Furlanetto} S.~R.,  {Zaldarriaga} M.,    {Hernquist} L.,  2004, ApJ, 613, 16

\bibitem[\protect\citeauthoryear{{Geil} \& {Wyithe}}{{Geil} \&
  {Wyithe}}{2008}]{geil2008}
{Geil} P.~M.,  {Wyithe} J.~S.~B.,  2008, MNRAS, 386, 1683

\bibitem[\protect\citeauthoryear{{Gnedin}}{{Gnedin}}{2000}]{gnedin2000}
{Gnedin} N.~Y.,  2000, ApJ, 542, 535

\bibitem[\protect\citeauthoryear{{Gnedin} \& {Fan}}{{Gnedin} \&
  {Fan}}{2006}]{gnedin2006}
{Gnedin} N.~Y.,  {Fan} X.,  2006, ApJ, 648, 1

\bibitem[\protect\citeauthoryear{{Gnedin} \& {Shaver}}{{Gnedin} \&
  {Shaver}}{2004}]{gnedin2004}
{Gnedin} N.~Y.,  {Shaver} P.~A.,  2004, ApJ, 608, 611

\bibitem[\protect\citeauthoryear{{Haiman}, {Abel} \& {Madau}}{{Haiman}
  et~al.}{2001}]{haiman2001}
{Haiman} Z.,  {Abel} T.,    {Madau} P.,  2001, ApJ, 551, 599

\bibitem[\protect\citeauthoryear{{Haiman} \& {Cen}}{{Haiman} \&
  {Cen}}{2002}]{haiman2002}
{Haiman} Z.,  {Cen} R.,  2002, ApJ, 578, 702

\bibitem[\protect\citeauthoryear{{Iliev}, {Mellema}, {Pen}, {Merz}, {Shapiro}
  \& {Alvarez}}{{Iliev} et~al.}{2006}]{iliev2006}
{Iliev} I.~T.,  {Mellema} G.,  {Pen} U.-L.,  {Merz} H.,  {Shapiro} P.~R.,
  {Alvarez} M.~A.,  2006, MNRAS, 369, 1625

\bibitem[\protect\citeauthoryear{{Kohler}, {Gnedin}, {Miralda-Escud{\'e}} \&
  {Shaver}}{{Kohler} et~al.}{2005}]{kohler2005}
{Kohler} K.,  {Gnedin} N.~Y.,  {Miralda-Escud{\'e}} J.,    {Shaver} P.~A.,
  2005, ApJ, 633, 552

\bibitem[\protect\citeauthoryear{{Kollmeier}, {Onken}, {Kochanek}, {Gould},
  {Weinberg}, {Dietrich}, {Cool}, {Dey}, {Eisenstein}, {Jannuzi}, {Le Floc'h}
  \& {Stern}}{{Kollmeier} et~al.}{2006}]{kollmeier2006}
{Kollmeier} J.~A.,  {Onken} C.~A.,  {Kochanek} C.~S.,  {Gould} A.,  {Weinberg}
  D.~H.,  {Dietrich} M.,  {Cool} R.,  {Dey} A.,  {Eisenstein} D.~J.,  {Jannuzi}
  B.~T.,  {Le Floc'h} E.,    {Stern} D.,  2006, ApJ, 648, 128

\bibitem[\protect\citeauthoryear{{Komatsu}, {Dunkley}, {Nolta}, {Bennett},
  {Gold}, {Hinshaw}, {Jarosik}, {Larson}, {Limon}, {Page}, {Spergel},
  {Halpern}, {Hill}, {Kogut}, {Meyer}, {Tucker}, {Weiland}, {Wollack} \&
  {Wright}}{{Komatsu} et~al.}{2008}]{komatsu2008}
{Komatsu} E.,  {Dunkley} J.,  {Nolta} M.~R.,  {Bennett} C.~L.,  {Gold} B.,
  {Hinshaw} G.,  {Jarosik} N.,  {Larson} D.,  {Limon} M.,  {Page} L.,
  {Spergel} D.~N.,  {Halpern} M.,  {Hill} R.~S.,  {Kogut} A.,  {Meyer} S.~S.,
  {Tucker} G.~S.,  {Weiland} J.~L.,  {Wollack} E.,    {Wright} E.~L.,  2008,
  arXiv:astro-ph/08030547

\bibitem[\protect\citeauthoryear{{Lidz}, {Zahn}, {McQuinn}, {Zaldarriaga} \&
  {Hernquist}}{{Lidz} et~al.}{2008}]{lidz2008}
{Lidz} A.,  {Zahn} O.,  {McQuinn} M.,  {Zaldarriaga} M.,    {Hernquist} L.,
  2008, ApJ, 680, 962

\bibitem[\protect\citeauthoryear{{Loeb}}{{Loeb}}{2006}]{loeb2006}
{Loeb} A.,  2006, ArXiv e-prints, 0603360

\bibitem[\protect\citeauthoryear{{Loeb} \& {Zaldarriaga}}{{Loeb} \&
  {Zaldarriaga}}{2004}]{loeb2004}
{Loeb} A.,  {Zaldarriaga} M.,  2004, Phys. Rev. Lett., 92, 211301

\bibitem[\protect\citeauthoryear{{Madau}, {Meiksin} \& {Rees}}{{Madau}
  et~al.}{1997}]{madau1997}
{Madau} P.,  {Meiksin} A.,    {Rees} M.~J.,  1997, ApJ, 475, 429

\bibitem[\protect\citeauthoryear{{Martini}}{{Martini}}{2004}]{martini2004}
{Martini} P.,  2004, in {Ho} L.~C.,  ed., Coevolution of Black Holes and
  Galaxies {QSO Lifetimes}.
pp 169--+

\bibitem[\protect\citeauthoryear{{Martini} \& {Weinberg}}{{Martini} \&
  {Weinberg}}{2001}]{martini2001}
{Martini} P.,  {Weinberg} D.~H.,  2001, ApJ, 547, 12

\bibitem[\protect\citeauthoryear{{McQuinn}, {Zahn}, {Zaldarriaga}, {Hernquist}
  \& {Furlanetto}}{{McQuinn} et~al.}{2006}]{mcquinn2006}
{McQuinn} M.,  {Zahn} O.,  {Zaldarriaga} M.,  {Hernquist} L.,    {Furlanetto}
  S.~R.,  2006, ApJ, 653, 815

\bibitem[\protect\citeauthoryear{{Meiksin}}{{Meiksin}}{2005}]{meiksin2005}
{Meiksin} A.,  2005, MNRAS, 356, 596

\bibitem[\protect\citeauthoryear{{Mesinger} \& {Furlanetto}}{{Mesinger} \&
  {Furlanetto}}{2007}]{mesinger2007}
{Mesinger} A.,  {Furlanetto} S.,  2007, ApJ, 669, 663

\bibitem[\protect\citeauthoryear{{Mo} \& {White}}{{Mo} \&
  {White}}{1996}]{mo1996}
{Mo} H.~J.,  {White} S.~D.~M.,  1996, MNRAS, 282, 347

\bibitem[\protect\citeauthoryear{{Morales} \& {Hewitt}}{{Morales} \&
  {Hewitt}}{2004}]{morales2004}
{Morales} M.~F.,  {Hewitt} J.,  2004, ApJ, 615, 7

\bibitem[\protect\citeauthoryear{{Shapiro}, {Giroux} \& {Babul}}{{Shapiro}
  et~al.}{1994}]{shapiro1994}
{Shapiro} P.~R.,  {Giroux} M.~L.,    {Babul} A.,  1994, ApJ, 427, 25

\bibitem[\protect\citeauthoryear{{Shaver}, {Windhorst}, {Madau} \& {de
  Bruyn}}{{Shaver} et~al.}{1999}]{shaver1999}
{Shaver} P.~A.,  {Windhorst} R.~A.,  {Madau} P.,    {de Bruyn} A.~G.,  1999,
  A\&A, 345, 380

\bibitem[\protect\citeauthoryear{{Shen}, {Strauss}, {Oguri}, {Hennawi}, {Fan},
  {Richards}, {Hall}, {Gunn}, {Schneider}, {Szalay}, {Thakar}, {Vanden Berk},
  {Anderson}, {Bahcall}, {Connolly} \& {Knapp}}{{Shen} et~al.}{2007}]{shen2007}
{Shen} Y.,  {Strauss} M.~A.,  {Oguri} M.,  {Hennawi} J.~F.,  {Fan} X.,
  {Richards} G.~T.,  {Hall} P.~B.,  {Gunn} J.~E.,  {Schneider} D.~P.,  {Szalay}
  A.~S.,  {Thakar} A.~R.,  {Vanden Berk} D.~E.,  {Anderson} S.~F.,  {Bahcall}
  N.~A.,  {Connolly} A.~J.,    {Knapp} G.~R.,  2007, AJ, 133, 2222

\bibitem[\protect\citeauthoryear{{Sheth}, {Mo} \& {Tormen}}{{Sheth}
  et~al.}{2001}]{sheth2001}
{Sheth} R.~K.,  {Mo} H.~J.,    {Tormen} G.,  2001, MNRAS, 323, 1

\bibitem[\protect\citeauthoryear{{Srbinovsky} \& {Wyithe}}{{Srbinovsky} \&
  {Wyithe}}{2007}]{srbin2007}
{Srbinovsky} J.~A.,  {Wyithe} J.~S.~B.,  2007, MNRAS, 374, 627

\bibitem[\protect\citeauthoryear{{Tozzi}, {Madau}, {Meiksin} \& {Rees}}{{Tozzi}
  et~al.}{2000}]{tozzi2000}
{Tozzi} P.,  {Madau} P.,  {Meiksin} A.,    {Rees} M.~J.,  2000, ApJ, 528, 597

\bibitem[\protect\citeauthoryear{{White}, {Becker}, {Fan} \& {Strauss}}{{White}
  et~al.}{2003}]{white2003}
{White} R.~L.,  {Becker} R.~H.,  {Fan} X.,    {Strauss} M.~A.,  2003, AJ, 126,
  1

\bibitem[\protect\citeauthoryear{{Wyithe} \& {Loeb}}{{Wyithe} \&
  {Loeb}}{2003}]{wl2003}
{Wyithe} J.~S.~B.,  {Loeb} A.,  2003, ApJ, 586, 693

\bibitem[\protect\citeauthoryear{{Wyithe} \& {Loeb}}{{Wyithe} \&
  {Loeb}}{2004}]{wl2004b}
{Wyithe} J.~S.~B.,  {Loeb} A.,  2004, ApJ, 610, 117

\bibitem[\protect\citeauthoryear{{Wyithe} \& {Loeb}}{{Wyithe} \&
  {Loeb}}{2007}]{wl2007}
{Wyithe} J.~S.~B.,  {Loeb} A.,  2007, MNRAS, 375, 1034

\bibitem[\protect\citeauthoryear{{Wyithe}, {Loeb} \& {Geil}}{{Wyithe}
  et~al.}{2008}]{wlg2008}
{Wyithe} J.~S.~B.,  {Loeb} A.,    {Geil} P.~M.,  2008, MNRAS, 383, 1195

\bibitem[\protect\citeauthoryear{{Wyithe} \& {Morales}}{{Wyithe} \&
  {Morales}}{2007}]{wm2007}
{Wyithe} J.~S.~B.,  {Morales} M.~F.,  2007, MNRAS, 379, 1647

\bibitem[\protect\citeauthoryear{{Wyithe} \& {Padmanabhan}}{{Wyithe} \&
  {Padmanabhan}}{2006}]{wp2006}
{Wyithe} J.~S.~B.,  {Padmanabhan} T.,  2006, MNRAS, 372, 1681

\end{thebibliography}

\end{document}